\documentclass[
aps,%
10pt,%
final,%
notitlepage,%
oneside,%
twocolumn,%
nobibnotes,%
nofootinbib,%
superscriptaddress,%
noshowpacs,%
centertags]%
{revtex4}
\begin{document}

\selectlanguage{english}
\title{Anisotropy of the Space Orientation of Radio Sources. I: The Catalog}
\author{\firstname{V.~R.}~\surname{Amirkhanyan}}

\affiliation{Sternberg Astronomical Institute, Universitetskii pr.
13, Moscow, 119992 Russia}

\received{January 21, 2009}%
\revised{May 20, 2009}%

\begin{abstract}
A catalog of extended extragalactic radio sources consisting of
10461 objects is compiled based on the list of radio sources of
the FIRST survey. A total of 1801 objects are identified with
galaxies and quasars of the SDSS survey and the Veron-Veron
catalog. The distribution of position angles of the axes of radio
sources from the catalog is determined, and the probability that
this distribution is equiprobable is shown to be{} less than
10$^{-7}$.  This result implies that at Z equal to or smaller than
0.5, the spatial orientation of the axes of radio sources is
anisotropic at a statistically significant level.

\end{abstract}

\maketitle
\section{INTRODUCTION}

The uniform morphology of galaxies make them excellent test bodies
for the investigations of the structure of space. The British
astronomer Brown was the first to study the orientation of
galaxies in 1930s \cite{1:Amirkhanyan1_n_en}. Brown~\cite{2:Amirkhanyan1_n_en}, Nilson~\cite{3:Amirkhanyan1_n_en},
\mbox{Lauberts~\cite{4:Amirkhanyan1_n_en},} and Karachentsev et al.~\cite{5:Amirkhanyan1_n_en}
compiled extensive catalogs providing the galaxy parameters that
are of most importance to such studies: position angles and axial
ratios. \mbox{Reinhardt~\cite{6:Amirkhanyan1_n_en},} Nilson~\cite{7:Amirkhanyan1_n_en},
Lauberts~\cite{4:Amirkhanyan1_n_en}, \mbox{Mandzhos \cite{8:Amirkhanyan1_n_en}}, and Parnovsky et
al.~\cite{9:Amirkhanyan1_n_en} analyzed these catalogs and showed convincingly that
the spatial orientation of galaxies is anisotropic on the scale
lengths of at least 200~Mpc. Extended extragalactic radio sources
can be seen to exhibit equally uniform and anisotropic structure.
Radio sources are very appealing objects for such studies, because
they should eventually allow to either prove or refute the results
based on an independent sample of galaxies and increase
substantially the volume of the space domain studied.
Amirkhanyan~\cite{10:Amirkhanyan1_n_en} tried to use such radio sources as
indicators of anisotropy, but failed to reach a conclusive result
because of a rather small size of the available homogeneous
sample, which consisted of  298 radio sources of the MG catalog.
It is evident that to obtain a statistically significant result, a
catalog of extended radio sources is required, comparable in size
with galaxy catalogs. However, no such catalogs are available to
date. Therefore, we decided, with the ultimate aim of analyzing
the spatial orientation of radio sources, to compile such a
catalog based on the list of objects of the FIRST
survey~\cite{11:Amirkhanyan1_n_en}. This catalog cannot be used directly since it
provides no information about the multiplicity of radio sources.
The catalog gives the parameters of individual components without
indicating eventual associations between them.

\begin{figure*}[tbp]
\setcaptionmargin{5mm} \onelinecaptionsfalse \captionstyle{normal}
\centerline{
\includegraphics[width=16cm, bb= 29 431 481 714, clip]{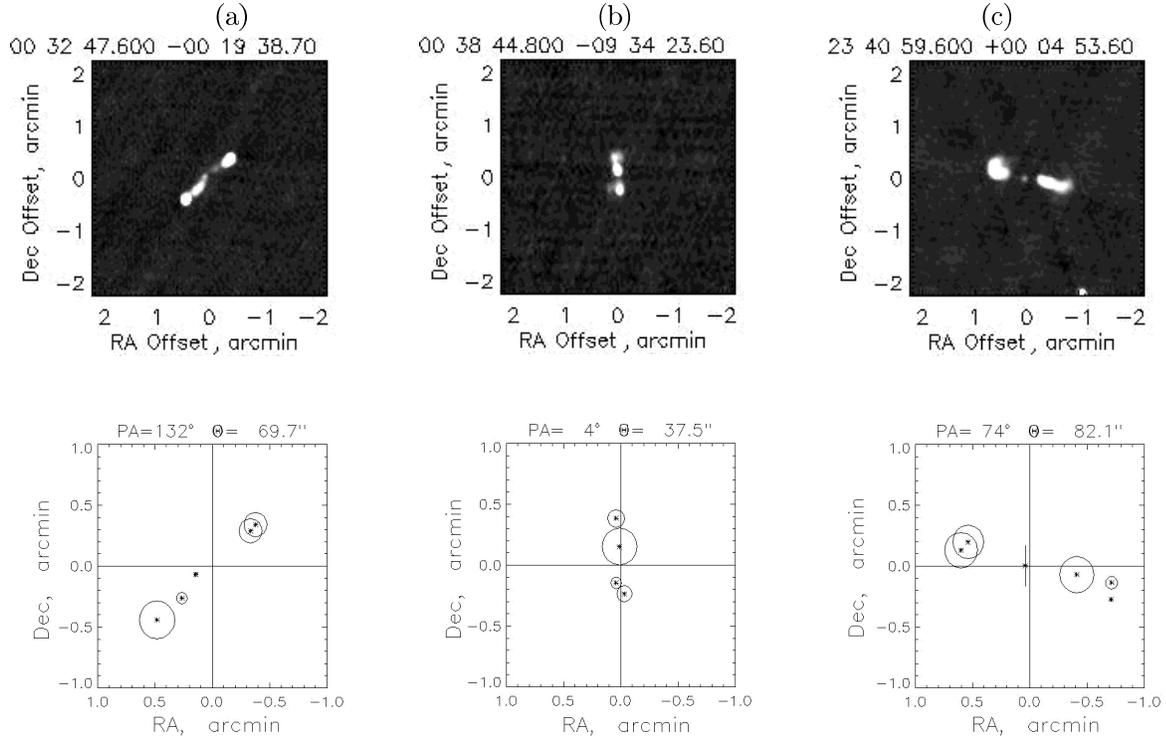}
} \caption{Images of radio sources in the  FIRST survey (the upper
line) and in this catalog (the lower line). The intersection of
the $\alpha$ and $\delta$ axes shown on the maps of the lower line
coincides with the average coordinates of the combined radio
source. The asterisks indicate the positions of the components,
and the diameters of the rings surrounding the asterisks are
proportional to the integrated fluxes of the components. The cross
indicates the location of the optical component  (the figure on
the right).} \label{fig_image:Amirkhanyan1_n_en}
\end{figure*}

\begin{figure*}[tbp]
\setcaptionmargin{5mm} \onelinecaptionstrue \captionstyle{normal}
\centerline{\includegraphics[width=9cm, bb= 61 179 547 546,
clip]{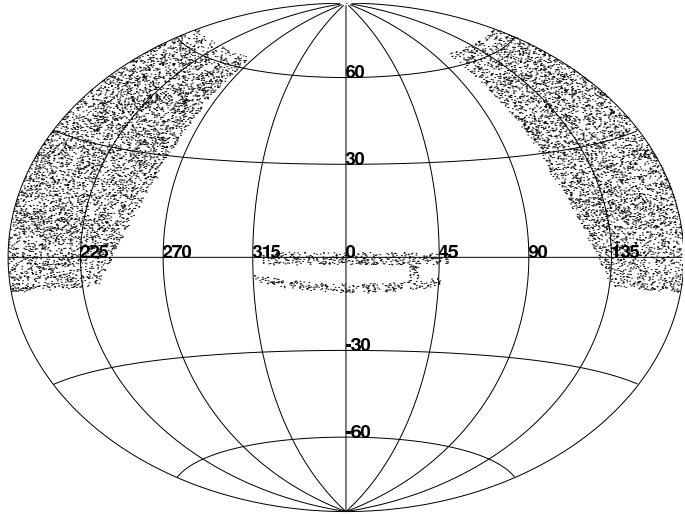}} \caption{Sky distribution of the
radio sources of the catalog.} \label{fig_sky:Amirkhanyan1_n_en}
\end{figure*}

\section{THE CATALOG}
\label{intro:Amirkhanyan1_n_en}
\subsection{Rules of Selection}
For the task to succeed, minimizing the number of false objects is
more important than finding all the true radio sources. Hence, the
selection rules embedded into the \textbf{first01} program that
produces the catalog, focused on the former rather than the latter
goal. The rules are as follows:

(1) each radio source must consist of at least three components;

(2) the integrated flux from each component must be greater than
or equal to 2.5 mJy;

(3) the root mean square distance of the components from the axis
of the radio source drawn optimally across the coordinates of the
components must not exceed  0.12 $\Theta $, where $\Theta$ is the
separation between the most distant components.

We use the cluster analysis methods to generate the list of
components that make up the radio source. Under these conditions,
if we assume that all the 440046 objects of the FIRST survey with
integrated fluxes equal to or greater than 2.5 mJy are mutually
unassociated, we should find no more than three false radio
sources with three components. The probability of finding a false
object with four or more components is negligible. Another
possible source of false objects are very extended radio sources,
where the separation between the groups of components exceeds the
clusterization radius of $60''$. In this case, the program may
mistake a group of components, if it meets the above rules of
selection, for an independent radio source. Such objects are rare
and most of them do not satisfy the rules of selection. The
program yielded a catalog consisting of 10461 radio sources.
Figure~\ref{fig_image:Amirkhanyan1_n_en} shows examples of objects combined and
their radio images in the FIRST survey. We performed such visual
control for several tens of radio sources and it revealed no
errors whatsoever. The program simultaneously identified the radio
sources combined with objects of the SDSS survey~ \cite{12:Amirkhanyan1_n_en} and
the Veron-Veron \mbox{Catalog \cite{13:Amirkhanyan1_n_en}.} If the separation
between the average coordinates of the radio source and optical
object was less than $30''$ and less than half the angular size of
radio source, the object was included into the catalog as a
possible optical component. We then determine the component of the
radio source that is nearest to the optical object, and compute
the separation between them. At the same, we compute the distance
from the optical object to the axis of the radio source. In
subsequent studies we considered the radio source to have an
optical identification if at least one of these separations was
less than $3''$. A total of  1801 sources meet this condition.

\subsection{Simulation of the Catalog}
Visual control is limited in scope and subjective in nature.
Therefore, to test the quality of the search program, we generated
several simulations of the FIRST survey. The
\textbf{sim{\_}first01} code generated the preset number of
extended radio sources in accordance with the given sky
distribution. The number of components in radio sources was set in
a random way from three to ten. To make the model as realistic as
possible, we chose the distributions of fluxes, angular sizes, and
scatter of components about the axis of the object to be close to
the corresponding distributions  in the real catalog. Position
angle is the most important parameter of radio sources for our
task, and therefore the simulation code allowed the form of the
given distribution of position angles to be changed so as to
compare it with the position angles of simulated objects. We
number this list as 1. After completing the simulation, the
program generated the combined list of  {\it the components} of
all simulated radio sources and wrote their coordinates and fluxes
into a file in the format of the FIRST catalog \mbox{(list
No.~2).} This file was then processed by the \textbf{first01}
program, which composed list No.~3 consisting of simulated
objects. A comparison of lists Nos.~1 and 3 for several
simulations showed that about 8.5\% of the objects of list No.~1
fail to make it into list No.~3, since the scatter of their
components about the axis exceeds the preset threshold
(\mbox{selection rule No.~3)}. Furthermore, about 4\%  of all
objects are discarded as the separation between the groups of
their components exceeds the clusterization radius. No false
objects lacking in list No.~1 have been found in list No.~3.
Reduction of simulated catalogs revealed a serious problem to
arise when one has to compute the position angle of the axis of a
radio source in case where this angle is close to 0 or 90 degrees.
This problem is easy to explain. Let $x_i $ and $y_i $ be the
coordinates of $N$ components with respect to the center of mass
of the radio source. The denominator or numerator of the formula
for the slope of the straight line drawn through the set of points
using the least squares method always contains $\sum\nolimits_{i =
1}^N {x_i^2 } $  or $\sum\nolimits_{i = 1}^N {y_i^2 } $. The
coordinates of the components always include measurement errors,
and therefore the above sums obey the $\chi ^2$ distribution, and
the probability that they should reduce to zero is negligible.
Hence the probability to correctly estimate the position angle
decreases as the real position angle approaches to 0 or 90 degrees
(depending on the method used to draw the line) and, consequently,
the number of objects of the histogram at these positions should
also steadily decrease. As a result, we obtain a distorted
distribution of position angles. To overcome these limitations,
the program determines each position angle via a two-stage
process. First, it uses standard formulas to find the straight
line with the lowest sum of squared distances from the components,
and computes its position angle $\phi$. The program then refines
the position angle by varying the slope of the line within
$\phi\pm 10\degr$, and seeks the real minimum of squared
distances. Numerical simulations showed that this computation
method eliminates the problem of finding the position angle: the
errors between the position angles of simulated (list No.~1) and
combined (list No.~3) objects do not depend on the angle value,
and the forms of the preset and measured position angle
distributions of the objects from the simulated catalog coincide.

\subsection{Format of the Catalog}
The final catalog of extended radio sources has the layout as
described in the Table.

\setcaptionmargin{5mm}
\onelinecaptionstrue
\captionstyle{nonumber}
\begin{longtable*}{c|c|c|c|c|c}
\caption{{\bf Table.} Extended radio sources}\\
\hline
 $\alpha_{2000}$ &  $\delta_{2000}$  &  \multicolumn{4}{c}{Characteristics of the source}  \\
\hline
(1)& (2)& (3)& (4)& (5)& (6)\\
\hline
\endfirsthead
\caption{{\bf Table.} (Contd.)}\\
\hline
 $\alpha_{2000}$ &   $\delta_{2000}$&  \multicolumn{4}{c}{Characteristics of the source} \\
\hline
(1)& (2)& (3)& (4)& (5)& (6)\\
\hline
\endhead
\hline
\endfoot

\hline
\endlastfoot

 129.550781   &  17.674994  &    97.0  &  74.2 &   27.8&     \\
   129.540726 &  17.675444  &   1.3 &    3.2 & 0.0 &  \\
   129.549530 &  17.675861  &   2.5  &   4.6 & 0.0&  \\
   129.551239 &  17.675165  &   3.1  &   5.5 & 0.0&   \\
   129.562271 &  17.673082  &   5.7   & 14.5  &0.0&  \\
\hline
   129.545731&   31.886414  &  82.6  &  25.2 &27.6&  \\
   129.542221 &  31.886194 &    3.8  &  10.2 &0.0&   \\
   129.546097 &  31.886360 &    7.6  &   7.8  &1.0&  \\
   129.546127&   31.886318&    17.7&  0.1699& 0.2&  --0.3\\
   129.550400&   31.887083&     1.9&     9.6& 0.0&   \\
\hline
   129.555511&   13.968682&   38.1&   137.2&  52.8&  \\
   129.543793&   13.953083&    21.9&    26.5&  0.0&  \\
   129.554718&   13.969610&     8.6&     8.8&  1.0&   \\
   129.554718&   13.969643&   -18.7&  2.0134&  0.1& 6.7 \\
   129.568069&   13.983055&    14.5&    17.5&  0.0&   \\
\hline
   129.567520&   26.378452&   1.2&    44.8&    115.5&  \\
   129.567352&   26.374390&    13.7&    10.9&  0.0&   \\
   129.567673&   26.376307&    19.2&    53.7&  0.0&   \\
   129.567749&   26.386833&    36.0&    51.0&  0.0&    \\
\hline
   129.580429&   40.770386&    14.7&    29.5&  18.5&   \\
   129.579269&   40.767113&     2.4&     7.3&  0.0&   \\
   129.580505&   40.770500&     1.9&     3.1&  0.0&   \\
   129.582016&   40.775028&     1.7&     8.1&  0.0&    \\
\hline
   129.586914&   17.208118&    73.0&    43.1& 90.5&   \\
   129.581558&   17.207390&     8.3&    33.1&  0.0&   \\
   129.585205&   17.206223&     2.5&    13.1&  0.0&   \\
   129.593689&   17.210306&    10.9&    44.3&  0.0&   \\
\hline
   129.595459&   12.498256&    80.2&    18.8& 179.5&   \\
   129.592773&   12.497806&    29.0&    35.6&  0.0&   \\
   129.595444&   12.498250&    44.7&    51.2&  1.0&   \\
   129.595474&   12.498255&   -19.1&  1.6294&  0.1& --0.0\\
   129.598053&   12.498694&    78.5&    92.7&  0.0&    \\
\hline
   129.610794&    0.017126&    21.9&    59.2&  29.7&    \\
   129.607422&    0.010861&     6.0&     8.9 &  0.0&   \\
   129.610001 &   0.012806&     6.2 &    7.9&   0.0&   \\
   129.614105&    0.025889 &   11.2&    12.9 & 0.0 & \\

\end{longtable*}

The data for each radio source is arranged in blocks. The first
line of each block describes the parameters of the object as a
whole:

\begin{itemize}
\item[(1)] average right ascension in degrees;

\item[(2)] average declination in degrees;

\item[(3)] position angle of the axis in degrees;

\item[(4)] angular size in arc seconds (we adopt the angular size
to be the separation between the mutually most distant components
of the radio source);

\item[(5)] integrated flux in mJy (the flux of a radio source is
the sum of the integrated fluxes of its components). The three
latter parameters are arranged into the column named
``Characteristics of the source''.
\end{itemize}


\begin{figure*}[tbp]
\setcaptionmargin{5mm} \onelinecaptionsfalse \captionstyle{normal}
\centerline{\includegraphics[width=9cm, bb= 31 252 562 567,
clip]{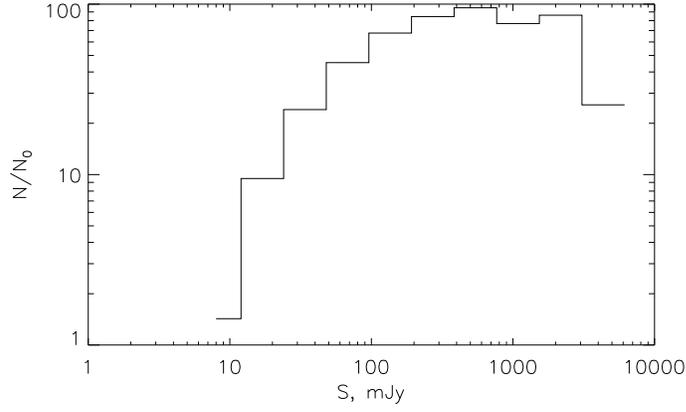}} \caption{Differential flux
distribution of the radio sources of the catalog normalized to the
statistics in isotropic Euclidean space.} \label{fig_NS:Amirkhanyan1_n_en}
\end{figure*}

\begin{figure*}[tbp]
\setcaptionmargin{5mm} \onelinecaptionstrue \captionstyle{normal}
\centerline{\includegraphics[width=9cm, bb= 23 251 551 737,clip]
{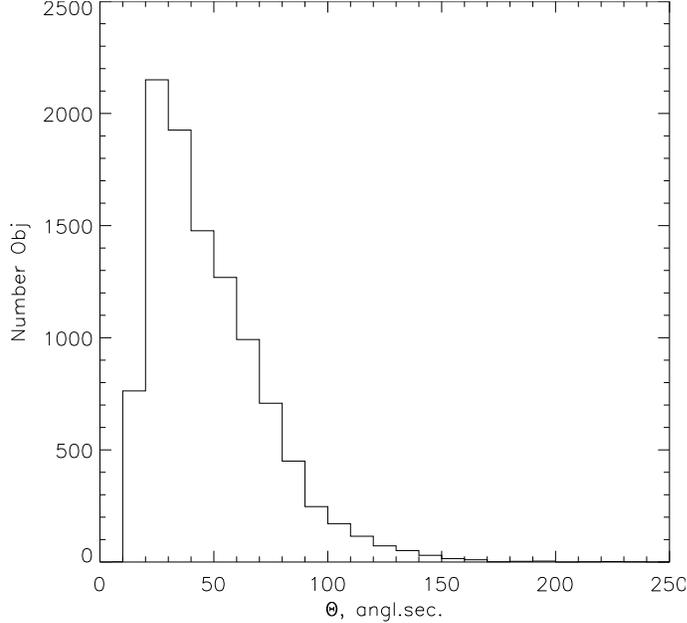}} \caption{Distribution of angular sizes of
radio sources.} \label{fig_NT:Amirkhanyan1_n_en}
\end{figure*}

\begin{figure*}[tbp]
\setcaptionmargin{5mm} \onelinecaptionsfalse \captionstyle{normal}
\centerline{\includegraphics[width=9cm, bb= 30 252 568 739,
clip]{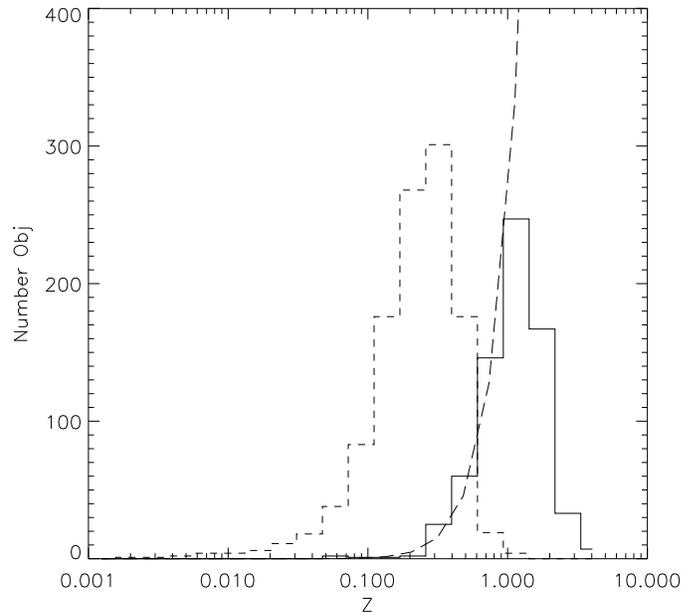}} \caption{Redshift distribution of
identified radio sources of the catalog. The short-dashed and
solid lines show the distributions for galaxies and quasars,
respectively. The long-dashed line shows the computed dependence
for the Einstein--de Sitter model.} \label{fig_NZ:Amirkhanyan1_n_en}
\end{figure*}

\begin{figure*}[tbp]
\setcaptionmargin{5mm} \onelinecaptionsfalse \captionstyle{normal}
\includegraphics[width=10cm, bb= 21 254 565 737, clip]{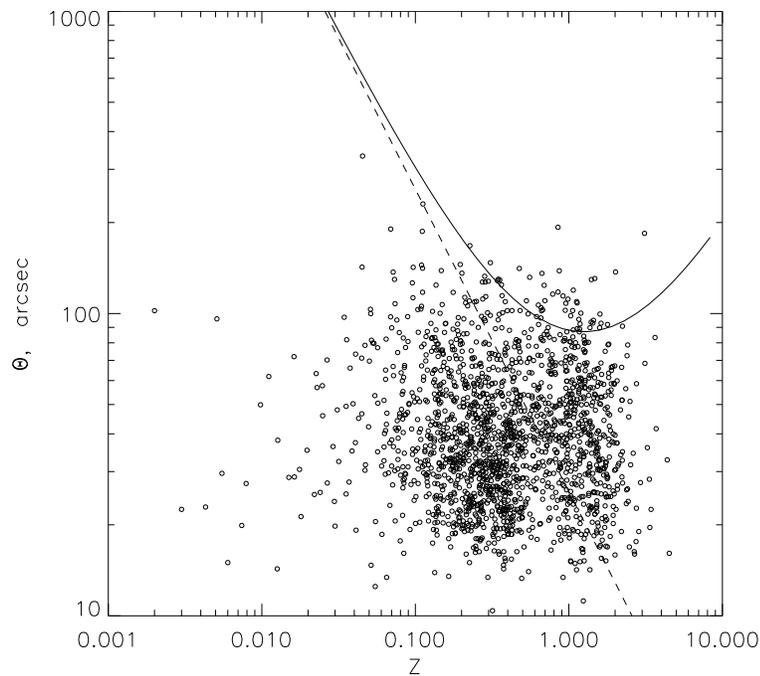}
 \caption{The ``angular size--redshift'' relation. Also shown here are the dependencies of radio sources with the linear
size of 500~kpc in the Euclidean (the dashed line) and
Einstein--de Sitter (the solid line) models.} \label{fig_TZ:Amirkhanyan1_n_en}
\end{figure*}

\begin{figure*}[tbp]
\setcaptionmargin{5mm} \onelinecaptionsfalse
\centerline{
\includegraphics[width=14cm, bb= 27 220 494 713,clip]{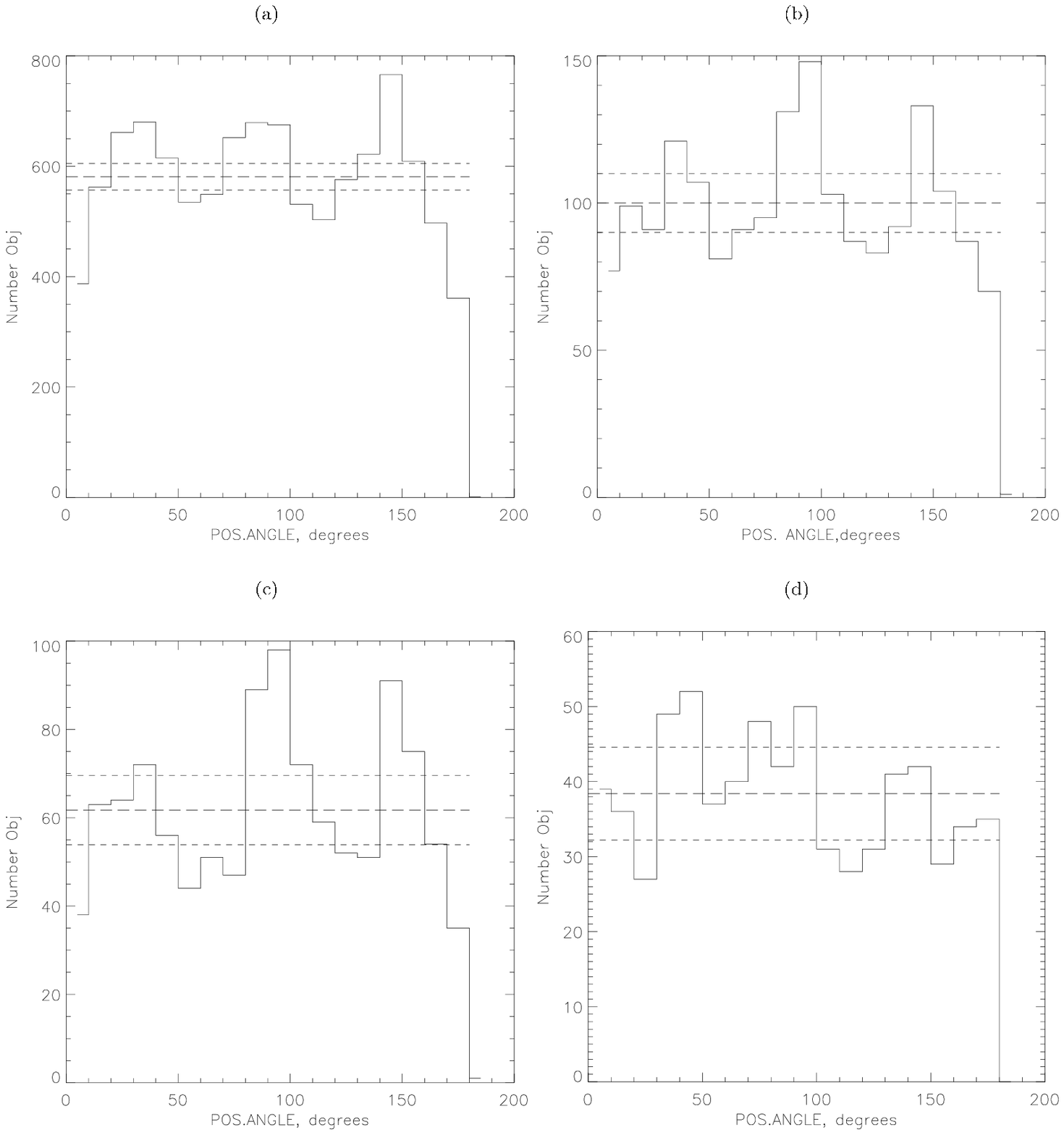}
} \captionstyle{normal} \caption{Distributions of position angles
of the radio sources of the catalog. The dashed lines show the
average level and the error interval under the assumption of
equiprobable distribution of angles. Shown are: the distribution
of the position angles of the radio sources of the catalog (a);
the distribution of the position angles of the radio sources of
the catalog with \mbox{known Z (b);} the distribution of the
position angles of the radio sources of the catalog with known  Z
and identified with \mbox{galaxies (c),} and the histogram of the
position angles of the radio sources of the catalog with known Z
and identified with \mbox{quasars (d).} }
\label{fig_PA:Amirkhanyan1_n_en}
\end{figure*}

Every subsequent line gives the data on every component that makes
up the radio source considered. These data are adopted from the
FIRST catalog:

\begin{itemize}

\item[(1)] right ascension in degrees;

\item[(2)] declination in degrees;

\item[(3)] flux of the unresolved component in mJy;

\item[(4)] integrated flux of the component in mJy;

\item[(5)] zero, if the component is not identified with any
optical object, and one if such identification exists. Unity
indicates that the next line contains the data on the optical
object:
\end{itemize}

\begin{itemize}

\item[(1)] right ascension in degrees;

\item[(2)] declination in degrees;

\item[(3)] V- or g-band magnitude.  Negative magnitude means that
the object is a quasar. If the absolute value of this parameter is
equal to unity it means that we could not find either V- or g-band
magnitude for the object;

\item[(4)] redshift;

\item[(5)] separation between the optical object and the
corresponding radio component identified by the program, in
arcsec;

\item[(6)] distance between the optical object and the axis of the
radio source in arcsec.
\end{itemize}

The catalog in the above format is available at the web-page of
the Special Astrophysical Observatory of the Russian Academy of
Sciences at: {\tt ftp://ftp.sao.ru/cifs/cats/cats/FIRST\_Amir/}
{\tt /f1912kak5.txt}.

\section{STATISTICS OF THE CATALOG}

Figure~\ref{fig_sky:Amirkhanyan1_n_en} shows the sky distribution of the radio
sources of the catalog, which coincides with the sky area of the
FIRST survey. Figure~\ref{fig_NS:Amirkhanyan1_n_en} shows the differential
distribution of the fluxes of the radio sources of the catalog
normalized to the ``Euclidean statistics''. Figure~\ref{fig_NT:Amirkhanyan1_n_en}
shows the distribution of angular sizes of the radio sources. The
resulting probability density of the distribution of angular sizes
of the radio sources of the catalog in the  $20'' - 200''$
interval can be approximated fairly well by the following
function:

$$
 P(\theta )d\theta = 0.025e^{ - 0.000247(\theta-3.3)^2}d\theta.
$$

Figure~\ref{fig_NZ:Amirkhanyan1_n_en} shows the distribution of the redshifts of
radio sources identified: galaxies (the short-dashed line) and
quasars (the solid line). This figure also shows the redshift
dependence of the number of objects in terms of the Einstein--de
Sitter model (the long-dashed line). It is evident that the
cessation of the increase and subsequent rapid decline of the
number of objects with increasing  Z is due to the instrumental
limitations of optical observations, and not to the actual
decrease of the density of these objects.  At small Z, the
computed dependence agrees fairly well with the experiment, and
this should be viewed as an evidence of absence of any significant
selection effects during the composition of the catalog. Of great
interest is the ``angular size--redshift'' relation
(Fig.~\ref{fig_TZ:Amirkhanyan1_n_en}). Here we also show the relations derived in
terms of the Euclidean (the dashed line) and Einstein--de Sitter
(the solid line) models for radio sources with the linear size of
500\,kpc. Legg~\cite{14:Amirkhanyan1_n_en} and Miley~\cite{15:Amirkhanyan1_n_en} constructed this
relation using objects of the 3CR catalog and showed that the
upper envelope of the plot in angular size can be described fairly
well in terms of the Euclidean model ($\theta \sim
\raise0.7ex\hbox{$1$} \!\mathord{\left/ {\vphantom {1
Z}}\right.\kern-\nulldelimiterspace}\!\lower0.7ex\hbox{$Z$})$. To
explain the discrepancy between this result and the main models of
space, the above authors suggested that linear sizes of radio
sources may evolve as \mbox{$D \sim (1 + Z)^{ - 1.5}$.}
\mbox{Amirkhanyan \cite{16:Amirkhanyan1_n_en}} showed that the observed upper
envelope of this plot can be explained by the selection effect due
to the limited sensitivity of the surveys and anisotropy of the
directivity pattern of the radio sources. The detection threshold
of the FIRST survey is three orders of magnitude lower than that
of the 3CR survey, and therefore the selection boundary should
have moved far upward in terms of redshift. It is evident from
Fig.\,6 that the upper limit of the new diagram agrees better with
the standard model than with the Euclidean model. However, the
parameters of the model are so far difficult to estimate even from
the data for 1801 objects.

\section{CONCLUSIONS}
Figure~\ref{fig_PA:Amirkhanyan1_n_en}a shows the distribution of position angles of
the axes of the radio sources. Its $\chi ^2$ value is $\chi
^2$=314.4, and the probability of it being isotropic is less than
10$^{-7}$. This result inevitably implies that the spatial
orientation of the axes of the radio sources is anisotropic. This
naturally brings up the question where this anisotropy is located.
The distribution of position angles of the axes of the radio
sources with known redshifts (Fig.7b) is similar to the
corresponding distribution for the entire sample. However, the
distributions of the position angles of the radio sources with
known redshifts identified with  1112 galaxies (Fig.7c) and 691
quasars (Fig.7d) differ appreciably from each other --- galaxies
exhibit a much stronger nonuniformity than quasars. Does this mean
that galaxies are the main contributors to the nonuniformity of
the distribution and anisotropy extends only out to Z $\sim $
0.5--1? We do not yet know the answer to this question. And our
second (cautious) conclusion is: the fact that the minimum of the
histogram lies near 0$\degr$ means that the axes of the radio
sources are mostly oriented in the direction of the equator.

\begin{acknowledgments}
The author is grateful to I.~D.~Karachentsev for fruitful
discussion of this work.
\end{acknowledgments}


\begin{thebibliography}{99}
\bibitem{1:Amirkhanyan1_n_en}
 F.~G.~Brawn,
\mnras~ \textbf{99}, 534 (1938).
\bibitem{2:Amirkhanyan1_n_en}
F.~G.~Brawn, \mnras~
\textbf{127}, 517 (1963).
\bibitem{3:Amirkhanyan1_n_en}
 P.~Nilson, Ann. Uppsala astron. Observ. \textbf{6}, 1 (1973).
\bibitem{4:Amirkhanyan1_n_en}
 A.~Lauberts, The ESO/Uppsala Survey of the ESO(B) Atlas. (1982).
\bibitem{5:Amirkhanyan1_n_en}
I.~D.~Karachentsev, V.~E.~Karachentseva, and S.~L.~Parnovsky,
Astron.
Nachr. \textbf{314}, 97 (1993).
\bibitem{6:Amirkhanyan1_n_en}
 M. Reinhardt, \mnras~ \textbf{156}, 151 (1972).
\bibitem{7:Amirkhanyan1_n_en}
 P.~Nilson, Rep. Uppsala astron. Observ. \textbf{3}, 1 (1974).
\bibitem{8:Amirkhanyan1_n_en}
 A.~V.~Mandzhos, A.~Ya.~Gregul', I.~Yu.~Izotova, and V.~V.~Tel'nyuk-Adamchuk,
Astrofizika \textbf{26}, 321 (1987).
\bibitem{9:Amirkhanyan1_n_en}
 S.~L.~Parnovsky, I.~D.~Karachentsev, and V.~E.~Karachentseva,
\mnras~ \textbf{268}, 665 (1994).
\bibitem{10:Amirkhanyan1_n_en}
 V.~R.~Amirkhanian, \bsao~ \textbf{37}, 119 (2000).
\bibitem{11:Amirkhanyan1_n_en}
 R.~L.~White, R.~H.~Becker, D.~J.~Helfand, and M.~D.~Gregg, \aj~
\textbf{475}, 479 (1997).
\bibitem{12:Amirkhanyan1_n_en}
{\tt http://cas.sdss.org/astro/en/tools/crossid/ \linebreak
/crossid.asp}

\bibitem{13:Amirkhanyan1_n_en}
M.~P.~Veron-Cetty and P.~Veron, \aaa~ \textbf{455},
773 (2006).
\bibitem{14:Amirkhanyan1_n_en}
 T.~H.~Legg, Nature \textbf{226}, 65 (1970).
\bibitem{15:Amirkhanyan1_n_en}
 G.~K.~Miley,
\mnras~ \textbf{152}, 477 (1971).
\bibitem{16:Amirkhanyan1_n_en}
 V.~R.~Amirkhanyan, Soobshchenija Spetsial'noi Astrofiz. Obs., \textbf{61}, 112 (1989).
\end{thebibliography}
\end{document}